\documentclass[reprint, superscriptaddress, amsmath, amssymb, aps, pra]{revtex4-1}
\usepackage{graphicx}
\usepackage{dcolumn}
\usepackage{bm}
\usepackage[dvipdfm,colorlinks,linkcolor=blue, urlcolor=blue, anchorcolor=blue, citecolor=blue]{hyperref}
\usepackage{color}
\usepackage{ulem}
\usepackage{amsmath,amsfonts,amssymb}
\usepackage{booktabs}
\usepackage{float}
\usepackage{graphicx}

\begin{document}
\title{Highly Efficient Creation and Detection of Ultracold Deeply-Bound Molecules via Chainwise Stimulated Raman Shortcut-to-Adiabatic Passage}\thanks{This manuscript has been submitted to Physical Review A.}

\author{Jiahui Zhang}%
\affiliation{School of Physics, East China University of Science and Technology, Shanghai 200237, China}

\author{Li Deng}%
\email[]{dengli@ecust.edu.cn}
\affiliation{School of Physics, East China University of Science and Technology, Shanghai 200237, China}

\author{Yueping Niu}
\affiliation{School of Physics, East China University of Science and Technology, Shanghai 200237, China}
\affiliation{Shanghai Frontiers Science Center of Optogenetic Techniques for Cell Metabolism, East China University of Science and Technology, Shanghai 200237, China}
\affiliation{Shanghai Engineering Research Center of Hierarchical Nanomaterials, East China University of Science and Technology, Shanghai 200237, China}

\author{Shangqing Gong}
\email[]{sqgong@ecust.edu.cn}
\affiliation{School of Physics, East China University of Science and Technology, Shanghai 200237, China}
\affiliation{Shanghai Frontiers Science Center of Optogenetic Techniques for Cell Metabolism, East China University of Science and Technology, Shanghai 200237, China}
\affiliation{Shanghai Engineering Research Center of Hierarchical Nanomaterials, East China University of Science and Technology, Shanghai 200237, China}

\begin{abstract}
Chainwise stimulated Raman adiabatic passage (C-STIRAP) in M-type molecular system
is a good alternative in creating ultracold deeply-bound molecules when the typical STIRAP in
$\Lambda$-type system does not work due to weak Frank-Condon factors between states.
However, its creation efficiency under the smooth evolution is generally low.
During the process,
the population in the intermediate states may decay out quickly and
the strong laser pulses may induce multi-photon processes.
In this paper, we find that shortcut-to-adiabatic (STA) passage fits very well in improving the performance
of the C-STIRAP.
Currently, related discussions on the so-called chainwise stimulated Raman shortcut-to-adiabatic passage (C-STIRSAP) are rare.
Here, we investigate this topic
under the adiabatic elimination.
Given a relation among the four
incident pulses,
it is quite interesting that
the M-type system can be generalized into an effective $\Lambda$-type structure with the simplest resonant coupling.
Consequently,
all possible methods of STA for three-state system can be borrowed.
We take the counter-diabatic driving and ``chosen path" method as instances to demonstrate our treatment on the molecular system.
Although the ``chosen path" method does not work well in real three-state system if there is strong decay in the excited state,
our C-STIRSAP protocol under both the two methods can create ultracold deeply-bound molecules with high efficiency
in the M-type system.
The evolution time is shortened without strong laser pulses and the robustness of STA is well preserved.
Finally, the detection of ultracold deeply-bound molecules is discussed.
\end{abstract}
\maketitle

\section{\label{sec:level1}Introduction}

The study of ultracold molecules is a rapidly developing area with fruitful research outcomes.
There is great interest
in their rich internal and motional quantum degrees of freedom, enabling the investigation of few-body collisional physics \cite{Blume}, ultracold chemistry \cite{Hutson}, precision measurements \cite{Ulmanis2012}, quantum computations \cite{PhysRevLett.88.067901} and quantum simulations \cite{Carr2009}, to name a few.
However, long-lived ensembles and full quantum state control are prerequisites to harness the potential of ultracold molecules.
Therefore, cooling interacting molecular gases deeply into the quantum degenerate regime is required, which remains an elusive challenge \cite{PhysRevA.105.040101}.

Currently, a versatile approach for this purpose involves the atom-by-atom assembly of ultracold molecules from pre-cooled atoms.
Initially, weakly-bound Feshbach molecules close to the atomic threshold are created via magnetically induced Feshbach association \cite{RevModPhys.78.483}.
Subsequently, the Feshbach molecules are quickly transferred into ultracold deeply-bound molecules, in which the population
lies in the lowest energy level of the electronic ground state.
Note only in this state can collisional stability be ensured.
As for other states, the molecules will undergo fast inelastic collisions that lead to trap loss.
Over the past few years, the well-known stimulated Raman adiabatic passage (STIRAP) \cite{RevModPhys.79.53}
has become the most widely used technique.
Under the Stokes and pump pulse with counter-intuitive time sequence, the population in the Feshbach state
can be completely transferred into the deeply-bound ground state without populating the intermediate excited state.
The STIRAP has been successfully demonstrated for species such as
homonuclear Rb$_2$, Cs$_2$, Sr$_2$ and Li$_2$ \cite{PhysRevLett.98.043201, PhysRevLett.96.050402, PhysRevLett.101.133005,Danzl1062, PhysRevLett.109.115302, PhysRevA.102.013310},
as well as heteronuclear KRb, RbCs, NaK, RbSr, NaRb, NaLi, RbHg, LiK,
and NaCs \cite{PhysRevLett.105.203001, Ospelkaus2008, Ni231, PhysRevA.94.022507,
PhysRevA.97.013405, PhysRevLett.122.253201,PhysRevLett.114.205302,
PhysRevLett.116.205303,PhysRevLett.119.143001, PhysRevLett.124.133203,PhysRevLett.126.123402, PhysRevLett.130.113002, PhysRevResearch.4.L022019, Warner_2023, PhysRevLett.125.083401,PhysRevA.97.013405}.

Nevertheless, the STIRAP is such an adiabatic passage that it has to satisfy the adiabatic condition for perfect transfer.
Therefore, the process usually evolves smoothly, rendering it intrinsically time consuming \cite{RevModPhys.79.53}.
In reality, the requirement usually results in imperfect transfer due to the presence of spontaneous emission and collision
\cite{PhysRevA.85.023629, PhysRevA.87.043631, Zhang_2021}.
Moreover, the violation of the evolution along the dark state also leads to imperfect
transfer if the adiabatic condition is not fulfilled \cite{PhysRevLett.113.205301}.

In order to overcome the problem associated with the adiabatic passage,
a technique which is called shortcut-to-adiabaticity (STA)
was proposed \cite{RevModPhys.91.045001}.
The STA can speed up the adiabatic process while maintaining consistent results.
Besides, the robustness of the adiabatic passage is also preserved even if the employed parameters do not meet the adiabatic condition.
Nowadays, the STA has been extensively studied and various methods have been raised for different purposes.
For instance, the counter-diabatic driving, or equivalently the transitionless quantum algorithm,
is a commonly used method \cite{Demirplak2003, Demirplak2005, doi:10.1063/1.2992152, Berry_2009, PhysRevLett.105.123003, Masuda2015}.
It usually requires an additional counter-diabatic field to directly couple the initial and target state to remove the non-adiabatic effect.
Except for this, there are also methods such as
invariant-based inverse engineering \cite{doi:10.1063/1.1664991, PhysRevA.83.062116, PhysRevA.86.033405, Ho:15, PhysRevA.96.023843},
multiple Schr\"{o}dinger dynamics \cite{PhysRevLett.109.100403, PhysRevA.93.052324}, superadiabatic dressed states \cite{PhysRevLett.116.230503, Zhou2017}
and three-mode parallel paths \cite{PhysRevA.100.043413, Wu:17, PhysRevApplied.18.014038}.
Based on these methods, the application of STA into the STIRAP
has been suggested \cite{PhysRevA.94.063411} and successfully demonstrated in experiment \cite{Du2016}.

When employing the STIRAP to create ultracold deeply-bound molecules,
it is also crucial to note that the excited state, which serves as a bridge for the transfer,
should be suitably identified.
The excited state should have favorable transition dipole moments and good Frank-Condon overlaps
with the vibrational wavefunctions of the Feshbach and ground molecular state \cite{Stwalley2004}.
Customarily, this requirement is uneasy to meet due to the large difference in the average extension
of the Feshbach and ground state.
If such an excited state can not be provided for an efficient transfer,
people can turn to using more intermediate states to form a M-type structure to overcome this technical challenge.
Under the chainwise coupling, the Feshbach state can also be transferred into the ground state
by using multiple Raman adiabatic passages \cite{Danzl2010, PhysRevA.78.021402, PhysRevA.82.011609}.
This method, stemming from the studies of STIRAP in atomic systems, is known as chainwise STIRAP
(C-STIRAP) \cite{PhysRevA.56.4929, PhysRevA.44.7442, PhysRevA.48.845, PhysRevA.32.2776}.
However, the creation efficiency of the C-STIRAP is generally low \cite{Danzl2010}.
The main reason is that the intermediate states are populated during the multiple-step process.
The spontaneous emission and collision then reduce the transfer efficiency under the long evolution time.
Although increasing the coupling between the intermediate states can suppress the population in these states, it may be problematic for molecules because multi-photon processes are expected to occur to break the process \cite{PhysRevLett.80.932}.

According to the unique properties of the STA mentioned above,
we find that the STA is quite suitable
to solve the problems that the C-STIRAP encountered in its applications.
Undoubtedly, it would be of interest if the STA can speed up the C-STIRAP to reduce the population loss
from the intermediate states by avoiding using strong coupling pulses.
Currently, related studies are quite few \cite{PhysRevA.102.023515} and this topic deserves more
investigations.

In this paper, we combine the STA with C-STIRAP in the M-type molecular system,
which we call the chainwise stimulated Raman
shortcut-to-adiabatic passage (C-STIRSAP), to study the creation of ultracold deeply-bound molecules with
high efficiency.
We first reduce the M-type system into an effective $\Lambda$-type system under the adiabatic elimination (AE).
By doing so, the major population decay in the excited states out of the system can be suppressed.
If we further set a requirement towards the relation among the four incident pulses,
the effective $\Lambda$-type system
will possess the simplest resonant couplings.
This generalized model permits us to borrow all possible methods of STA from three-state system into
the M-type system.
We take the counter-diabatic driving and ``chosen path" method as examples
to demonstrate the feasibility of the C-STIRSAP.
Note in real three-state system, the ``chosen path" method can not realize perfect transfer
if the excited state has a strong decay.
However,
numerical calculations show here that our protocol under the two methods can
both create ultracold deeply-bound molecules with nearly perfect efficiency
in the M-type system.
Moreover, the process is speeded up without strong laser pulses and the robustness of STA is still
maintained.
Since the STA usually requires accurate
control of the incident pulses, the requirement of the relation among the pulses in our protocol
actually does not add extra experimental complexity.
Finally, we discuss the detection of ultracold deeply-bound molecules.

\begin{figure}[b]
\centering{\includegraphics[width=6cm]{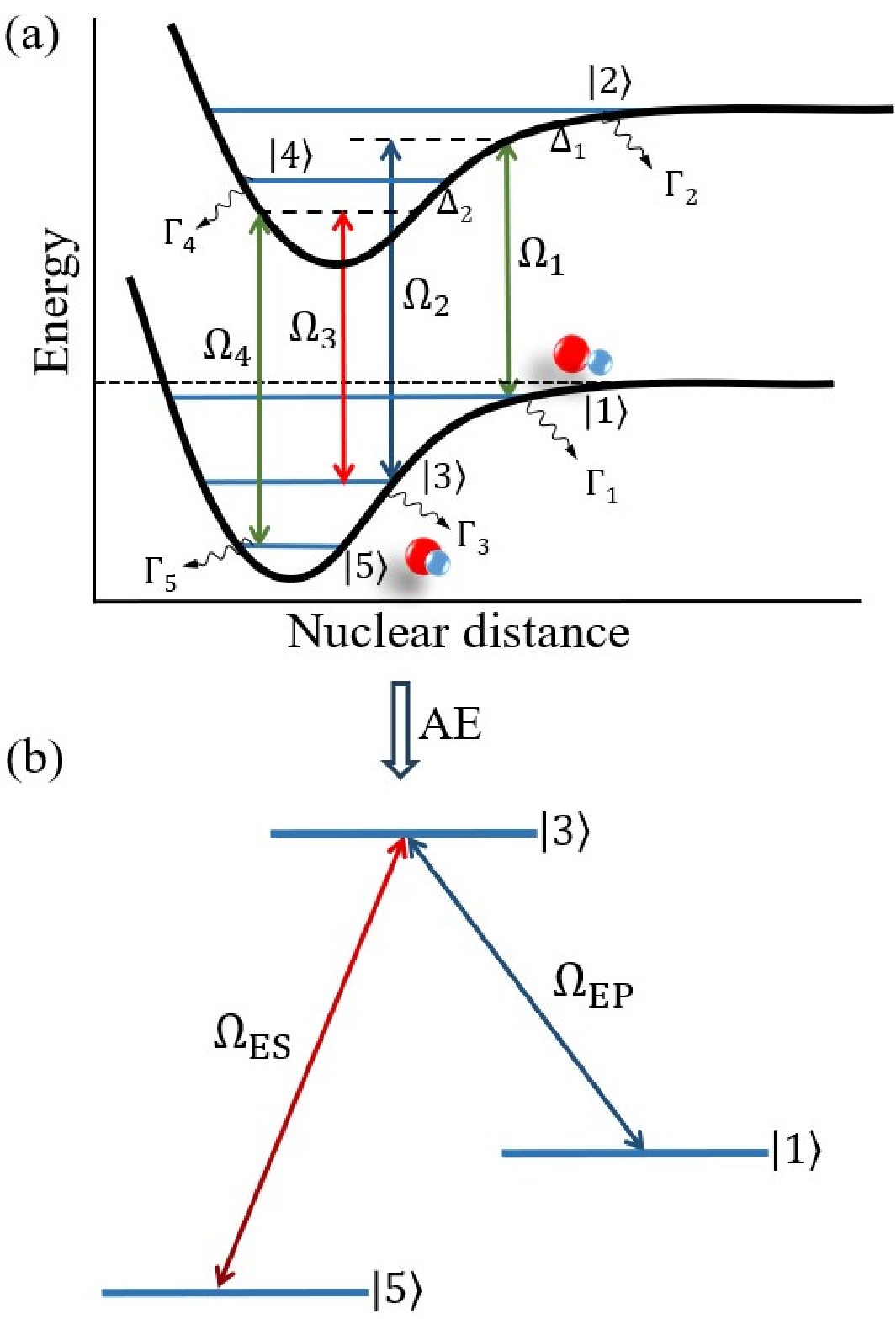}}
\caption{(a) Schematic diagram of the M-type molecular system with chainwise coupling.
(b) Effective $\Lambda$-type system with resonant couplings $\Omega_{\rm{EP}}$ and $\Omega_{\rm{ES}}$
after the AE.}
\label{fig1}
\end{figure}

\section{\label{sec:level2}generalization of the model}

We start by considering the M-type molecular structure with chainwise coupling as illustrated in Fig.~\ref{fig1}(a).
Here,
$|j\rangle (j=1, 3, 5)$
is the vibrational state of the ground electronic molecular state.
To be more accurate, $|1\rangle$, $|3\rangle$ and $|5\rangle$
are the Feshbach state, an intermediate state
and the deeply-bound ground state, respectively.
$|j\rangle (j=2, 4)$
is the vibrational state of an excited electronic molecular state.
As a bridge of the transfer, the two excited states are assumed to have good Franck-Condon factors with the three
ground states.
The coupling between states is presented by the time-dependent Rabi frequency $\Omega_{j} (j=1,2,3,4)$
in this figure.
Without loss of generality, all the states have decays $\Gamma_{j} (j=1,2,3,4,5)$ out of the molecular system
due to spontaneous emissions and inelastic collisions.
Nevertheless, we temporarily ignore them in the theoretical analysis in order to clarify the underlying physics.

The evolution of the molecular system is governed by the time-dependent Schr\"{o}dinger equation
\begin{eqnarray}\label{1}
i\hbar\displaystyle\frac{\partial}{\partial t} |\Phi(t)\rangle=H(t) |\Phi(t)\rangle.
\end{eqnarray}

\noindent
In this equation, $|\Phi(t)\rangle$ is the molecular wave function
with the expanding form
$| \Phi(t)\rangle=\sum_{j} a_{j}(t)| j\rangle (j=1,2,3,4,5)$,
where $a_{j}(t) (j=1,2,3,4,5) $ represents the probability amplitude of the corresponding state.
The two $\Lambda$-type substructures
$| 1\rangle\leftrightarrow |2 \rangle \leftrightarrow |3\rangle$ and
$| 3\rangle \leftrightarrow |4 \rangle\leftrightarrow |5\rangle$
in the molecular system are supposed to satisfy the two-photon resonance condition,
as required by the C-STIRAP technique.
Under the rotating wave approximation, Eq.~(\ref{1}) can be written as following
\begin{eqnarray}\label{2}
i\frac{d}{dt}
\left(
\begin{array}{ccccc}
a_1\\
a_2\\
a_3\\
a_4\\
a_5\\
\end{array}
\right)
=
\left(
\begin{array}{ccccc}
0&\Omega_{1}&0&0&0\\
\Omega_{1}&\Delta_1&\Omega_{2}&0&0\\
0&\Omega_{2}&0&\Omega_{3}&0\\
0&0&\Omega_{3}&\Delta_2&\Omega_{4}\\
0&0&0&\Omega_{4}&0\\
\end{array}
\right)
\left(
\begin{array}{ccccc}
a_1\\
a_2\\
a_3\\
a_4\\
a_5\\
\end{array}
\right).
\end{eqnarray}

\noindent
in which $\Delta_1$ and $\Delta_2$ are the single-photon detunings of the corresponding transitions.
We can clearly see from this equation that the Rabi frequency
always connects one of the ground
states with one of the excited states to form the chainwise coupling.

In order to apply the STA into the C-STIRAP, we are going to treat the above Hamiltonian in another way around,
which is different from that in Ref.~\cite{PhysRevA.102.023515}.
We first consider the use of
AE condition \cite{PhysRevA.82.011609} under the assumption that the two
detunings $\Delta_1$ and $\Delta_2$ are large, which is
\begin{equation}\label{3}
|\Delta_{1}|  \gg \sqrt{\Omega_{1}^2+\Omega_{2}^2}
\end{equation}

\noindent
and
\begin{equation}\label{4}
|\Delta_{2}|  \gg \sqrt{\Omega_{3}^2+\Omega_{4}^2},
\end{equation}

\noindent
respectively.
For simplicity, we assume that $\Delta_1=\Delta_2=\Delta$.
After the AE, the population in
the excited states is minimized and the decay out of the system can be greatly suppressed.
By setting $\frac{d}{dt}a_{2}=\frac{d}{dt}a_{4}=0$,
the molecular system can be reduced to an effective $\Lambda$-type structure
which includes only
ground states $|j\rangle (j=1, 3, 5)$
as shown in Fig.~\ref{fig1}(b).
The Schr\"{o}dinger equation is simplified as
\begin{eqnarray}\label{5}
i\frac{d}{dt}
\left(
\begin{array}{ccc}
a_1\\
a_3\\
a_5\\
\end{array}
\right)
=
\left(
\begin{array}{ccc}
\beta_1&\Omega_{\rm{EP}}&0\\
\Omega_{\rm{EP}}&\beta_3&\Omega_{\rm{ES}}\\
0&\Omega_{\rm{ES}}&\beta_5\\
\end{array}
\right)
\left(
\begin{array}{ccc}
a_1\\
a_3\\
a_5\\
\end{array}
\right).
\end{eqnarray}

\noindent
In this equation, the $3 \times 3$ matrix is the effective Halmiltonian with
the three diagonal elements being defined as
$\beta_1=-\frac{\Omega_{1}^2}{\Delta}$,
$\beta_3=-\frac{\Omega_{2}^2+\Omega_{3}^2}{\Delta}$ and
$\beta_5=-\frac{\Omega_{4}^2}{\Delta}$, respectively.
$\Omega_{\rm{EP}}$ and $\Omega_{\rm{ES}}$ can be considered as Rabi frequencies
induced by an effective pump and Stokes pulse and they are expressed as
$\Omega_{\rm{EP}}=-\frac{\Omega_{1}\Omega_{2}}{\Delta}$
and
$\Omega_{\rm{ES}}=-\frac{\Omega_{3}\Omega_{4}}{\Delta}$.
We can find that
the direct couplings between the ground and excited states now disappear and there are only
effective couplings between the
ground states $| 1\rangle \leftrightarrow | 3\rangle$ and
$|3\rangle \leftrightarrow | 5\rangle$.

We can continue to simplify the analysis if the three diagonal elements are equal to each other,
which is $\beta_1=\beta_3=\beta_5=\beta$.
This requires that the four Rabi frequencies $\Omega_{j} (j=1,2,3,4)$ should satisfy the following relation
\begin{eqnarray}\label{6}
\Omega_{1}=\Omega_{4}=\sqrt{\Omega_{2}^2+\Omega_{3}^2}.
\end{eqnarray}

\noindent
By further setting $a_j =c_{j}e^{-i\beta t}(j=1, 3, 5)$, Eq.~(\ref{5}) can be rewritten as
\begin{equation}\label{7}
\begin{split}
i\frac{d}{dt}
\left(
\begin{array}{ccc}
c_1\\
c_3\\
c_5\\
\end{array}
\right)
=
\left(
\begin{array}{ccc}
0&\Omega_{\rm{EP}}&0\\
\Omega_{\rm{EP}}&0&\Omega_{\rm{ES}}\\
0&\Omega_{\rm{ES}}&0\\
\end{array}
\right)
\left(
\begin{array}{ccc}
c_1\\
c_3\\
c_5\\
\end{array}
\right).
\end{split}
\end{equation}

\noindent
The two parameters $\Omega_{\rm{EP}}$ and $\Omega_{\rm{ES}}$
are changed to be
\begin{equation}\label{8}
\Omega_{\rm{EP}}=-\frac{\Omega_{2} \sqrt{\Omega_{2}^2+\Omega_{3}^2}}{\Delta}
\end{equation}

\noindent
and
\begin{equation}\label{9}
\Omega_{\rm{ES}}=-\frac{\Omega_{3} \sqrt{\Omega_{2}^2+\Omega_{3}^2}}{\Delta}
\end{equation}

\noindent
respectively.
Now the effective pump and Stokes pulse are resonant with the transitions $| 1\rangle \leftrightarrow | 3\rangle$ and
$|3\rangle \leftrightarrow | 5\rangle$.

Although we can reduce the M-type molecular system into the $\Lambda$-type system with
the simplest resonant coupling as shown in Fig.~\ref{fig1}(b),
still we can not directly employ the standard STIRAP.
Under the STIRAP,
the $\Lambda$-type system is expected to evolve adiabatically along
the dark state of the effective Hamiltonian in Eq.~(\ref{7}), which is
\begin{equation}\label{10}
| \Phi_{0}\rangle= \cos\theta |1 \rangle -\sin\theta |5\rangle
\end{equation}

\noindent
with the mixing angle being defined as $\tan\theta=\Omega_{\rm{EP}}/\Omega_{\rm{ES}}$.
Nevertheless, we need to satisfy the adiabatic condition $\sqrt{\Omega_{\rm{EP}}^2+\Omega_{\rm{ES}}^2} \gg \dot{\theta}$
\cite{RevModPhys.79.53}.
The condition requires using strong laser pulses and it conflicts with the AE condition; see Eqs.~(\ref{3}) and~(\ref{4}).
Therefore, the standard STIRAP is inappropriate for efficient creation of ultracold deeply-bound molecules here.
It is of importance if the STA can be applied here to avoid the use of strong pulses.

\section{\label{sec:level3}the C-STIRSAP}

In this section, we are going to carry out the STA combined with the C-STIRAP in the M-type
molecular system.
We call the technique the chainwise stimulated Raman shortcut-to-adiabatic passage (C-STIRSAP).
Based on the generalized three-state model in the above section, theoretically all kinds of methods
in the frame of STA can be borrowed here.
In the following, we will take the counter-diabatic driving, which is a widely used method, and the ``chosen path" method
as examples to illustrate the feasibility of the C-STIRSAP.

\subsection{counter-diabatic driving}

The basic idea of counter-diabatic driving
in a three-state $\Lambda$-type system
requires an auxiliary field to directly couple the two lower states
as shown in Fig.~\ref{fig1}(b)
to eliminate the non-adiabatic couplings and
make the adiabatic evolution possible.
Following the steps introduced in Ref.~\cite{PhysRevLett.105.123003}, we can obtain the expression
of the auxiliary field
\begin{equation}\label{11}
\Omega_{\rm{cd}}=\frac{\dot{\Omega}_2{\Omega}_3-\dot{\Omega}_3{\Omega}_2}{{\Omega}^2_2+{\Omega}^2_3},
\end{equation}

\noindent
in which
the overdot means the time derivative of the parameter.
Note that the auxiliary field now couples the Feshbach state
$|1\rangle$ and the deeply-bound
ground state $|5\rangle$.
Assisted by the counter-diabatic field,
the simplified $\Lambda$-type model
governed by Eq.~(\ref{7}) now becomes
\begin{equation}\label{12}
\begin{split}
i\frac{d}{dt}
\left(
\begin{array}{ccc}
c_1\\
c_3\\
c_5\\
\end{array}
\right)
=
\left(
\begin{array}{ccc}
0&\Omega_{\rm{EP}}&i\Omega_{\rm{cd}}\\
\Omega_{\rm{EP}}&0&\Omega_{\rm{ES}}\\
-i\Omega_{\rm{cd}}&\Omega_{\rm{ES}}&0\\
\end{array}
\right)
\left(
\begin{array}{ccc}
c_1\\
c_3\\
c_5\\
\end{array}
\right).
\end{split}
\end{equation}

\noindent
Turning back to the real M-type molecular system as shown in Fig.~\ref{fig1}(a),
the description of Eq.~(\ref{2}) with the auxiliary field is now modified to be
\begin{eqnarray}\label{13}
i\frac{d}{dt}
\left(
\begin{array}{ccccc}
a_1\\
a_2\\
a_3\\
a_4\\
a_5\\
\end{array}
\right)
=
\left(
\begin{array}{ccccc}
0&\Omega_{1}&0&0&i\Omega_{\rm{cd}}\\
\Omega_{1}&\Delta&\Omega_{2}&0&0\\
0&\Omega_{2}&0&\Omega_{3}&0\\
0&0&\Omega_{3}&\Delta&\Omega_{4}\\
-i\Omega_{\rm{cd}}&0&0&\Omega_{4}&0\\
\end{array}
\right)
\left(
\begin{array}{ccccc}
a_1\\
a_2\\
a_3\\
a_4\\
a_5\\
\end{array}
\right).
\end{eqnarray}

\noindent
Note that $\Omega_{\rm{cd}}$ still couples the states $|1\rangle$ and $|5\rangle$.
The above Schr\"{o}dinger equation after the modification
implies that we obtain a five-state counter-diabatic driving from
that of the three-state system.
The scheme is then supposed to achieve our goal that
it can speed up the process and preserve the adiabaticity without fulfilling
the adiabatic condition.
In order to verify the C-STIRSAP using the counter-diabatic driving
does work in the M-type molecular system,
we are going to study the evolution of the M-type system by numerical calculations.
Since the molecules suffer from unavoidable spontaneous emissions and collisions,
the decays from states displayed in Fig.~\ref{fig1}(a) should be taken into account.
Therefore, we are going to employ the Liouville-von Neumann equation
\begin{eqnarray}\label{14}
\displaystyle\frac{d\rho}{dt}=-i [\tilde{H}, \rho]-\Gamma\rho,
\end{eqnarray}

\noindent
for the calculations.
In this equation, the modified Hamiltonian $\tilde{H}$ is the $5 \times 5$ matrix in Eq.~(\ref{13}).
$\rho$ is the density matrix operator and $\Gamma\rho$ is the operator describing the decaying
processes. It is also a $5 \times 5$ matrix with its element defined as
$(\Gamma\rho)_{mn}=\frac{\Gamma_m+\Gamma_n}{2}\rho_{mn}(m,n=1,2,3,4,5)$.

We take the parameters of molecule $^{87}$Rb$_{2}$ in the calculations \cite{PhysRevA.78.021402}.
The population of the system is assumed to be initially in
the Feshbach state.
As for the decay rates,
the two excited states
$|2\rangle$ and $|4\rangle$
have the shortest lifetime and their decay rates are
chosen as $\Gamma_2=\Gamma_4=30{\rm{MHz}}$.
The decay rates of the Feshbach state $|1\rangle$ and the intermediate state
$|3\rangle$ are set to be
$\Gamma_1=0.01{\rm{MHz}}$ and $\Gamma_3 =0.06{\rm{MHz}}$, respectively.
The deeply-bound ground state $|5\rangle$ has quite a long lifetime compared with the
evolution time of the system.
Therefore, we can safely ignore the decay from this state, which is $\Gamma_5=0$.

The two pulses with Rabi frequencies $\Omega_2$ and $\Omega_3$ are partially overlapped and their envelopes are Gaussian shaped,
which are
\begin{eqnarray}\label{15}
\Omega_{\rm{2}}=\Omega_{0}\exp[-\frac{(t-t_{\rm{f}}/2-\tau)^2}{\sigma^2}]
\end{eqnarray}

\noindent
and
\begin{eqnarray}\label{16}
\Omega_{\rm{3}}=\Omega_{0}\exp[-\frac{(t-t_{\rm{f}}/2+\tau)^2}{\sigma^2}],
\end{eqnarray}

\noindent
respectively.
Here, $\Omega_{0}$ is the amplitude of the Rabi frequency.
$\sigma$ is the pulse duration. $2\tau$ is the time delay between the two pulses and
$t_{\rm{f}}$ is the evolution time (or operation time) of the system.
These parameters are chosen as $\Omega_0 =30\pi{\rm{MHz}}$,
$t_{\rm{f}}=1000 \rm{ns}$, $\sigma=t_{\rm{f}}/6$ and $\tau=t_{\rm{f}}/10$.
For the other two Rabi frequencies $\Omega_{\rm{1}}$ and $\Omega_{\rm{4}}$ and the
auxiliary field $\Omega_{\rm{cd}}$, they are chosen accordingly with respect to
Eqs.~(\ref{6}) and~(\ref{11}).
Note that the operation time is short and the adiabaticity is broken down in this case.
The detuning in the calculations is set to be $\Delta=2\pi{\rm{GHz}}$,
which is much larger than those Rabi frequencies to meet the requirement of AE.

\begin{figure}[t]
\centering{\includegraphics[width=8cm]{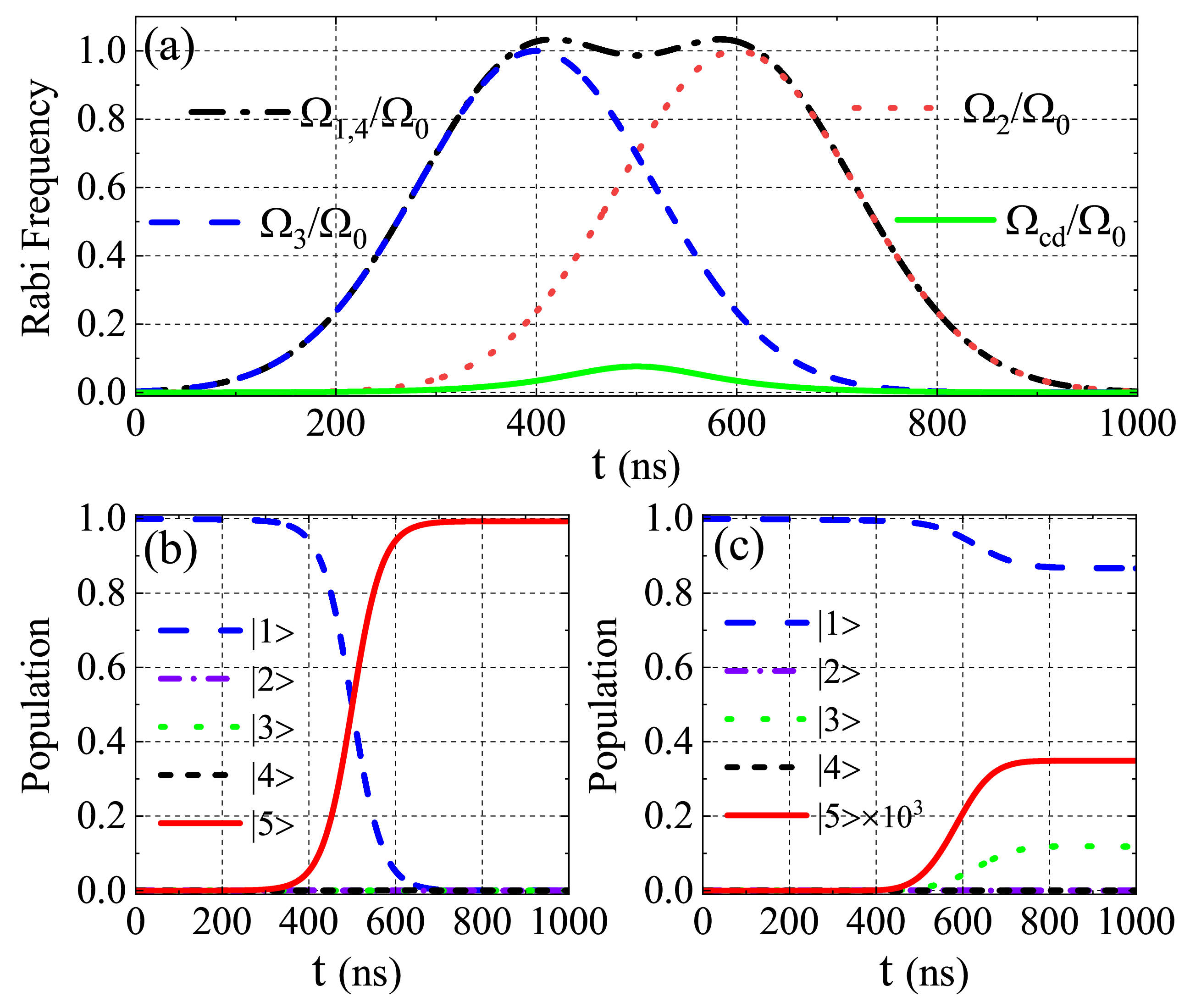}}
\caption{(a) Time sequence of the four Rabi frequencies $\Omega_j (j=1,2,3,4)$ together with the
auxiliary field $\Omega_{\rm{cd}}$.
(b) Population evolution under the C-STIRSAP by using the counter-diabatic driving while
(c) is that under the C-STIRAP without considering the auxiliary field.}
\label{fig2}
\end{figure}

We first present the evolution of the M-type molecular system under the C-STIRSAP with counter-diabatic
driving.
The time sequence of the four Rabi frequencies and the auxiliary field is shown in
Fig.~\ref{fig2}(a).
The corresponding population transfer in the system is displayed in Fig.~\ref{fig2}(b).
As a comparison, we also present the population evolution of the system under the C-STIRAP,
which can be seen in Fig.~\ref{fig2}(c).
The calculation of C-STIRAP is carried out by simply excluding the auxiliary field
in the modified Hamiltonian in Eq.~(\ref{13}) while
all the other parameters are kept unchanged.

We can observe from Fig.~\ref{fig2}(b) that we obtain nearly complete population transfer from the Feshbach state
$|1 \rangle$ to the ground
state $|5 \rangle$ without populating the intermediate state
$|3 \rangle$,
indicating the creation of ultracold deeply-bound molecules with
high efficiency.
The population of the excited states,
$|2 \rangle$ and $|4 \rangle$,
is negligible in presence of the AE.
On the contrary,
the population transfer from the Feshbach state $|1 \rangle$ to the ground state
$|5 \rangle$ hardly occurs in Fig.~\ref{fig2}(c), although
the intermediate state $|3 \rangle$ is partly populated.
The population in the excited states
$|2 \rangle$ and $|4 \rangle$ is still negligible
due to the AE.
Clearly,
the C-SRTRSAP with counter-diabatic driving can speed up the complete population transfer
without using strong pulses.

\begin{figure}[t]
\centering{\includegraphics[width=8cm]{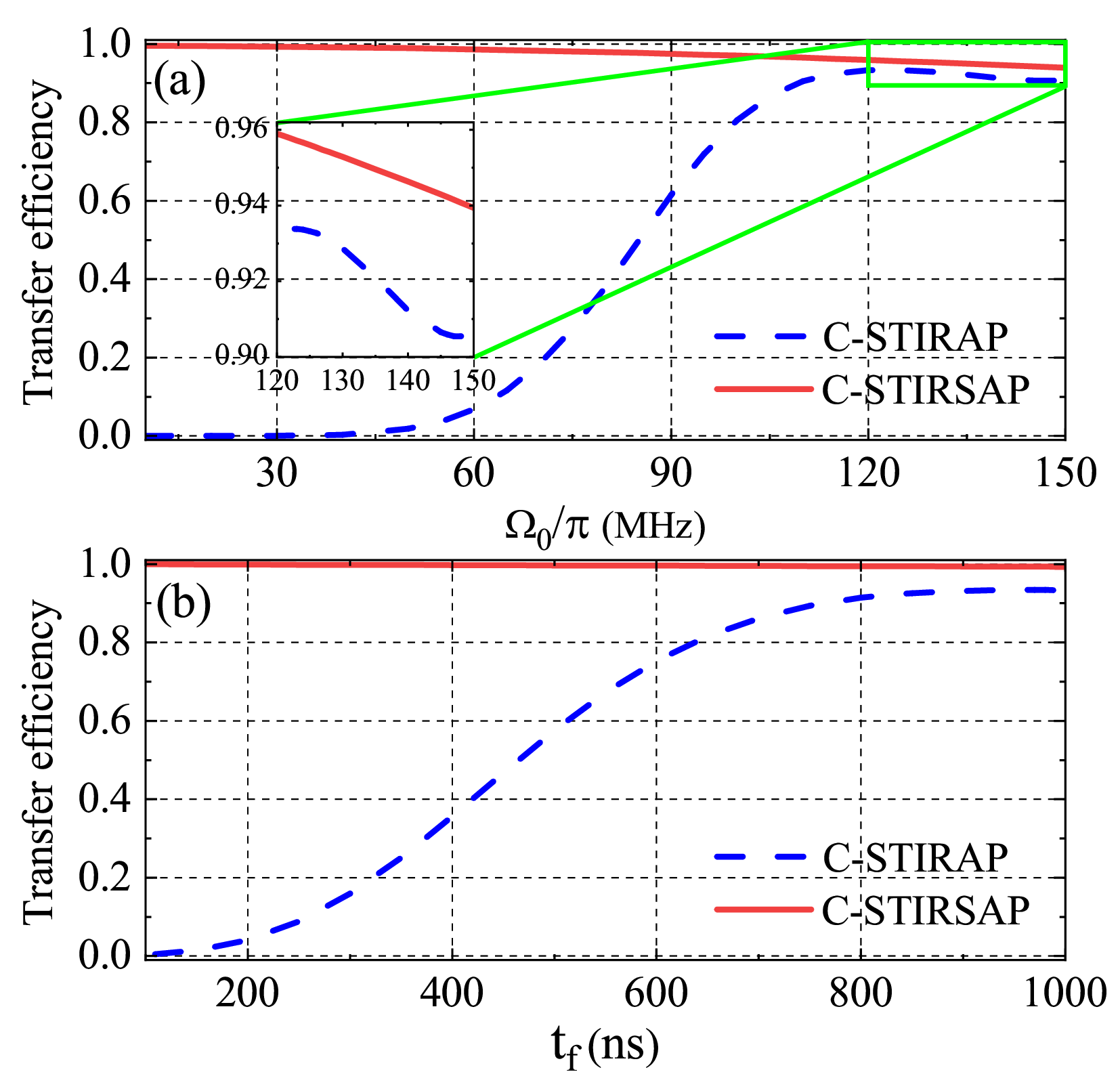}}
\caption{Transfer efficiency as functions of (a)
the amplitude of Rabi frequency $\Omega_{0}$ and (b)
the operation time $t_{\rm{f}}$ under the C-STIRSAP and C-STIRAP, respectively.
Note that in (b) $\Omega_0$ is chosen as $30\pi{\rm{MHz}}$ for the C-STIRSAP while it is $125\pi{\rm{MHz}}$
for the C-STIRAP. All the other parameters are the same as those in Fig.~\ref{fig2}.}
\label{fig3}
\end{figure}

Further calculations can show more properties of the C-STIRSAP with counter-diabatic driving.
In Fig.~\ref{fig3}(a), we study the transfer efficiency
from the Feshbach state $|1 \rangle$ to the ground state
$|5 \rangle$
as a function of the amplitude of Rabi frequency $\Omega_{0}$
under the C-STIRSAP and C-STIRAP, respectively.
All the other parameters are kept the same as those in Fig.~\ref{fig2}.
We can roughly see from Fig.~\ref{fig3}(a) that the transfer efficiency under the
C-STIRSAP is always higher than that under the C-STIRAP.
To be more precisely,
nearly perfect transfer efficiency can always be obtained under the C-STIRSAP.
The transfer efficiency slightly decreases with the increase of Rabi frequency $\Omega_{0}$.
The reason is the AE condition is gradually destroyed while the detuning
$\Delta$ is fixed unchanged,
resulting in the population in other states.
As for the
transfer efficiency under the C-STIRAP, it behaves differently.
Initially, it increases rapidly with the increase of Rabi frequency $\Omega_{0}$.
The efficiency reaches its maximum which is about
$94\%$ when the amplitude of the Rabi frequency is around $\Omega_{0}=125\pi{\rm{MHz}}$.
This is because the adiabaticity of the C-STIRAP is approximately satisfied
with the increase of the Rabi frequency and it
starts to dominate the process.
If we continue increasing the Rabi frequency, the efficiency starts to decrease,
which is also due to the breakdown of the AE condition.

In Fig.~\ref{fig3}(b),
we investigate the transfer efficiency
as a function of the operation time $t_{\rm{f}}$
under the C-STIRSAP and C-STIRAP, respectively.
Note the amplitude of Rabi frequency is chosen as
$\Omega_0 =30\pi{\rm{MHz}}$ for the C-STIRSAP,
while we choose the optimal value of $\Omega_0 =125\pi{\rm{MHz}}$ for the C-STIRAP,
which corresponds to the maximum
transfer efficiency in Fig.~\ref{fig3}(a).
The other parameters are kept unchanged as those in Fig.~\ref{fig2}.
We can see from Fig.~\ref{fig3}(b) that perfect transfer efficiency can always be obtained
regardless the choice of the operation time $t_{\rm{f}}$ under the C-STIRSAP.
In contrast, the transfer efficiency increases slowly with the increase of the operation time
$t_{\rm{f}}$ under the C-STIRAP, although the Rabi frequency $\Omega_0$ has been optimized.
Clearly, the C-STIRAP is time-consuming in order to obtain high transfer efficiency.

The above calculations demonstrate that the C-STIRSAP with counter-diabatic driving works
very well in
efficiently creating
ultracold deeply-bound molecules.
Besides,
the unique properties of STA are still preserved.
The operation time is shortened while the process is robust against the
fluctuations of parameters even
if the adiabatic condition is not fulfilled.
Nevertheless, we note that sometimes the direct coupling between states
$|1 \rangle$ and
$|5 \rangle$ by the auxiliary field is infeasible.
There may be possibilities that the transition is difficult to achieve due to different spin characters
or weak Franck-Condon factor between the states.
Therefore, alternative methods other than the direct application of counter-diabatic pulse are necessary.
In the following, we are going to take the ``chosen path" method to give another solution
to combine the STA with the C-STIRAP.

\subsection{``chosen path" method}

Unlike the counter-diabatic driving, the three-state ``chosen path" method accelerates the adiabatic process
by adding
an extra pair of fields into the Rabi frequencies $\Omega_{\rm{EP}}$ and $\Omega_{\rm{ES}}$ \cite{PhysRevA.100.043413}.
This idea can also be found in other methods such as invariant-based inverse engineering
\cite{doi:10.1063/1.1664991, PhysRevA.83.062116, PhysRevA.86.033405, Ho:15, PhysRevA.96.023843},
multiple Schr\"{o}dinger dynamics \cite{PhysRevLett.109.100403, PhysRevA.93.052324} and
superadiabatic dressed states \cite{PhysRevLett.116.230503, Zhou2017}.
Although these methods have different mathematical treatments, they are strongly related since
they possess similar underlying physics
(see recent reviews \cite{RevModPhys.91.045001}, \cite{ TORRONTEGUI2013117}
and \cite{doi:10.1080/23746149.2021.1894978}).
The two Rabi frequencies are then modified and the Schr\"{o}dinger equation describing the reduced $\Lambda$-type model now
has the form
\begin{equation}\label{17}
\begin{split}
i\frac{d}{dt}
\left(
\begin{array}{ccc}
c_1\\
c_3\\
c_5\\
\end{array}
\right)
=
\left(
\begin{array}{ccc}
0&\tilde{\Omega}_{\rm{EP}}&0\\
\tilde{\Omega}_{\rm{EP}}&0&\tilde{\Omega}_{\rm{ES}}\\
0&\tilde{\Omega}_{\rm{ES}}&0
\end{array}
\right)
\left(
\begin{array}{ccc}
c_1\\
c_3\\
c_5\\
\end{array}
\right).
\end{split}
\end{equation}

The aim of ``chosen path" method is that the simplified three-state system in
Fig.~\ref{fig1}(b) can evolve along one of its dressed states,
which is
\begin{eqnarray}\label{18}
| \Phi'_{0}\rangle =
&&\cos\mu(t)\cos\theta(t)|1 \rangle + i\sin\mu(t)|3 \rangle \nonumber \\
&&+\cos\mu(t)\sin\theta(t)|5 \rangle.
\end{eqnarray}

\noindent
Meanwhile, this dressed state $| \Phi'_{0}\rangle $ is designed to be decoupled with the other
two dressed states\cite{Wu:17}.
In order to transfer the population from the Feshbach state
$|1 \rangle$ to the deeply-bound ground
state $|5 \rangle$, the two time-dependent parameters $\mu(t)$ and $\theta(t)$ should satisfy the boundary condition,
which is
$\mu(0)=\mu(t_{\rm{f}})=0$ and
$\theta(0)= 0$ and $\theta(t_{\rm{f}})=\pi/2$.
Following the steps in Ref.~\cite{Wu:17},
the two parameters can be defined as
\begin{equation}\label{19}
\mu(t)=\frac{\gamma}{2}\left[1-\cos(\frac{2\pi t}{t_{\rm{f}}})\right]
\end{equation}

\noindent
and
\begin{equation}\label{20}
\theta(t)=\frac{\pi t}{2t_{\rm{f}}}-\frac{1}{3}\sin(\frac{2\pi t}{t_{\rm{f}}})+\frac{1}{24}\sin(\frac{4\pi t}{t_{\rm{f}}}),
\end{equation}

\noindent
respectively.
Then the corresponding two modified Rabi frequencies $\tilde{\Omega}_{\rm{EP}}$ and $\tilde{\Omega}_{\rm{ES}}$
with the following expressions
\begin{equation}\label{21}
\tilde{\Omega}_{\rm{EP}}=-\dot{\theta}(t)\sin\theta(t)\cot\mu(t)-\dot{\mu}(t)\cos\theta(t)
\end{equation}

\noindent
and
\begin{equation}\label{22}
\tilde{\Omega}_{\rm{ES}}=\dot{\theta}(t)\cos\theta(t)\cot\mu(t)-\dot{\mu}(t)\sin\theta(t)
\end{equation}

\noindent
can guarantee the evolution along the dressed state $| \Phi'_{0}\rangle$.
In Eq.~(\ref{19}), the parameter $\gamma$ should not be equal to zero.
Otherwise,
$\tilde{\Omega}_{\rm{EP}}$ and $\tilde{\Omega}_{\rm{ES}}$ will be infinitely large.

Now we go back to the M-type molecular system and try to design
the modified Rabi frequencies $\tilde{\Omega}_{j}(j=1,2,3,4)$ by comparing the Hamiltonians in Eqs.~(\ref{17})
and~(\ref{7}).
Similar to Eqs.~(\ref{8}) and~(\ref{9}), we can impose that
\begin{eqnarray}\label{23}
\tilde{\Omega}_{\rm{EP}}=-\frac{\tilde{\Omega}_{2}\sqrt{\tilde{\Omega}_{2}^2+\tilde{\Omega}_{3}^2}}{\tilde{\Delta}}
\end{eqnarray}

\noindent
and
\begin{eqnarray}\label{24}
\tilde{\Omega}_{\rm{ES}}=-\frac{\tilde{\Omega}_{3}\sqrt{\tilde{\Omega}_{2}^2+\tilde{\Omega}_{3}^2}}{\tilde{\Delta}},
\end{eqnarray}

\noindent
respectively.
Therefore, we
inversely derive the modified Rabi frequencies $\tilde{\Omega}_{2}$ and $\tilde{\Omega}_{3}$
as following
\begin{eqnarray}\label{25}
\tilde{\Omega}_{2}=\tilde{\Omega}_{\rm{EP}} \left( \frac{\tilde{\Delta}^2}
{\tilde{\Omega}^2_{\rm{EP}}+\tilde{\Omega}^2_{\rm{ES}}} \right) ^{\frac{\rm{1}}{\rm{4}}}
\end{eqnarray}

\noindent
and
\begin{eqnarray}\label{26}
\tilde{\Omega}_{3}=\tilde{\Omega}_{\rm{ES}} \left( \frac{\tilde{\Delta}^2}
{\tilde{\Omega}^2_{\rm{EP}}+\tilde{\Omega}^2_{\rm{ES}}} \right) ^\frac{1}{4}.
\end{eqnarray}

\noindent
We can also impose the modified Rabi frequencies $\tilde{\Omega}_{j}(j=1,2,3,4)$ satisfy
the similar relation as that in Eq.~(\ref{6}),
which is $\tilde{\Omega}_{1, 4}=\sqrt{\tilde{\Omega}_{2}^2+\tilde{\Omega}_{3}^2}$.
Then we obtain
\begin{eqnarray}\label{27}
\tilde{\Omega}_{1, 4}=\left[\tilde{\Delta}^2(\tilde{\Omega}^2_{\rm{EP}}+\tilde{\Omega}^2_{\rm{ES}})\right]^{\frac{1}{4}}.
\end{eqnarray}

\noindent
In order to make sure the reduced $\Lambda$-type system with modified effective pump and
Stokes pulse can be transformed back to the M-type system with modified Rabi frequencies $\tilde{\Omega}_{j}(j=1,2,3,4)$,
we should satisfy the AE condition or equivalently $\tilde{\Delta}\gg \tilde{\Omega}_{j}(j=1,2,3,4)$.
Here it is reasonable to assume $\tilde{\Delta}=\Delta$ since the detuning $\Delta$ is on the order of $\rm{GHz}$ while
the modified Rabi frequencies $\tilde{\Omega}_{j}(j=1,2,3,4)$ are on the order of $\rm{MHz}$ (see parameters in Fig.~\ref{fig5}).
Therefore, the evolution of the M-type molecular system
after modifications will follow
\begin{eqnarray}\label{28}
i\frac{d}{dt}
\left(
\begin{array}{ccccc}
a_1\\
a_2\\
a_3\\
a_4\\
a_5\\
\end{array}
\right)
=
\left(
\begin{array}{ccccc}
0&\tilde{\Omega}_{1}&0&0&0\\
\tilde{\Omega}_{1}&\Delta&\tilde{\Omega}_{2}&0&0\\
0&\tilde{\Omega}_{2}&0&\tilde{\Omega}_{3}&0\\
0&0&\tilde{\Omega}_{3}&\Delta&\tilde{\Omega}_{4}\\
0&0&0&\tilde{\Omega}_{4}&0\\
\end{array}
\right)
\left(
\begin{array}{ccccc}
a_1\\
a_2\\
a_3\\
a_4\\
a_5\\
\end{array}
\right).
\end{eqnarray}

The above equation implies that we obtain a five-state ``chosen path" method.
We can continue to
carry out numerical calculations by using the full Hamiltonian in Eq.~(\ref{28})
together with Eq.~(\ref{14}) to verify if
this method also works well in speeding up the C-STIRAP without maintaining the adiabatic
condition.

\begin{figure}[t]
\centering{\includegraphics[width=8.5cm]{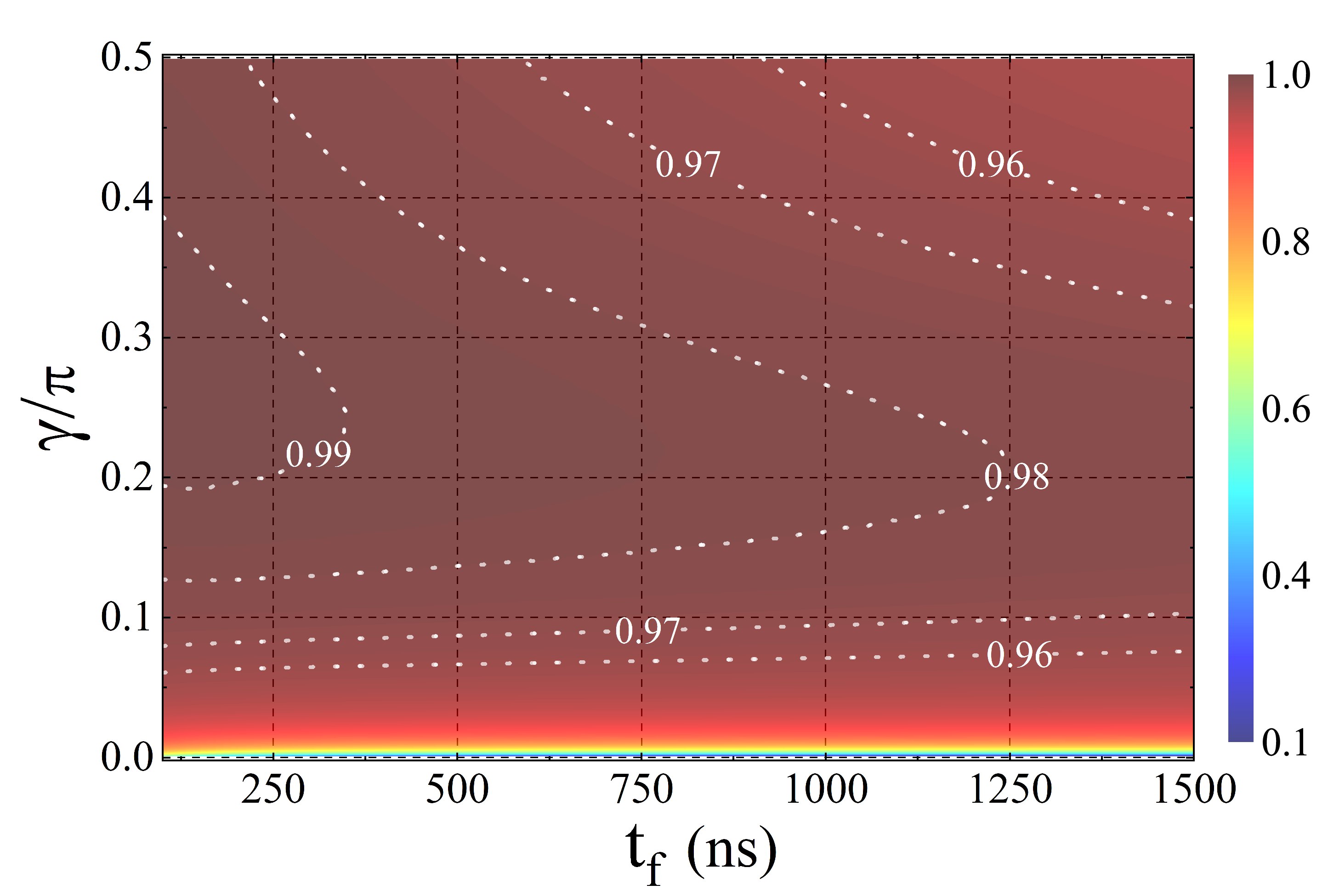}}
\caption{Transfer efficiency as functions of parameter $\gamma$ and operation time $t_{\rm{f}}$.
Note the starting value of $\gamma$ is $0.001\pi$ since it can not be equal to zero.}
\label{fig4}
\end{figure}

The ``chosen path" method in a real $\Lambda$-type system \cite{Wu:17} implies that
if there is no decay from the excited state,
perfect transfer efficiency
can always be achieved regardless the choice of the parameter $\gamma$ and operation time $t_{\rm{f}}$.
However, its efficiency can be very low if the strong decay in the excited state is considered.
In Fig.~\ref{fig4}, we also study the transfer efficiency as functions of the two parameters to see whether
the robustness is still preserved in M-type system.
In the calculations, the decay rates of molecule $^{87}$Rb$_{2}$ are also taken into account.
The detuning is fixed at $\Delta=2\pi{\rm{GHz}}$ and
the parameter $\gamma$ starts at the value of $0.001\pi$ since it can not be equal to zero.

We can see from Fig.~\ref{fig4} that even in presence of strong decay from the intermediate states,
nearly perfect transfer efficiency can be obtained in the M-type system within a wide range of the chosen
parameters $\gamma$ and $t_{\rm{f}}$,
which is quite opposite to the case when the ``chosen path" method is applied in a real three-state system.
The reason is the main population loss from the two excited states,
$|2 \rangle$ and $|4 \rangle$, is firstly
suppressed by the AE.
Secondly, the excited state $|3 \rangle$ in the generalized $\Lambda$-type system is actually a
ground state with small decay rate.
Therefore, the transient population in this state does not greatly influence the transfer efficiency.

In Fig.~\ref{fig4}, we can also see that the property of the ``chosen path" method is almost
preserved in M-type system.
The transfer efficiency is robust against the operation time $t_{\rm{f}}$ when the parameter $\gamma$ is fixed.
Slight difference only lies in choosing the value of parameter $\gamma$ when the operation time $t_{\rm{f}}$ is fixed,
which is $\gamma$ should not be too small in order for
the perfect transfer efficiency.
The reason is $\tilde{\Omega}_{\rm{EP}}$ and $\tilde{\Omega}_{\rm{ES}}$ tend to be very large at a small value
of $\gamma$.
Consequently, the AE condition is gradually broken down at the large value of
$\tilde{\Omega}_{j} (j=1,2,3,4)$.
The excited states
$|2 \rangle$ and $|4 \rangle$ will be populated and
the population will decay quickly out of the system, resulting in the decrease of the transfer efficiency.

Although Fig.~\ref{fig4} demonstrates that highly efficient creation of ultracold deeply-bound molecules
under the C-STIRSAP with ``chosen path"
method can be completed in a very short time,
it does not provide another information which we also concern very much.
That is how strong the four modified Rabi frequencies $\tilde{\Omega}_{j} (j=1,2,3,4)$ will be under this protocol.
Calculations reveal that the variations of the four Rabi frequencies $\tilde{\Omega}_{j} (j=1,2,3,4)$
under different choice of the operation time
$t_{\rm{f}}$ are similar when the parameter $\gamma$ is fixed.
Therefore, we can only investigate
the amplitudes of $\tilde{\Omega}_{1}$ and $\tilde{\Omega}_{4}$ as a function of
$t_{\rm{f}}$.
This is because the amplitudes of $\tilde{\Omega}_{2}$ and $\tilde{\Omega}_{3}$ are always smaller than
those of $\tilde{\Omega}_{1}$ and $\tilde{\Omega}_{4}$
(see the equation $\tilde{\Omega}_{1, 4}=\sqrt{\tilde{\Omega}_{2}^2+\tilde{\Omega}_{3}^2}$).
Here we set $\gamma=0.3\pi$ as an example and the result is shown by the red solid line in Fig.~\ref{fig5}(a).

It is obvious from Fig.~\ref{fig5}(a) that there is a trade-off between the pulse intensities and operation
time.
Shorter operation time of the C-STIRSAP requires stronger pulse intensities.
On one hand, the pulse intensities can not be too strong in order to maintain the AE condition.
On the other hand, the operation time can not be too long in order to avoid the population loss through
the intermediate states.
Hence, the two parameters should be chosen reasonably according to the actual situation.
Besides, we point out that the pulse intensities can be further decreased by using
a smaller detuning $\Delta$.
This is because the four Rabi frequencies $\tilde{\Omega}_{j} (j=1,2,3,4)$ are proportional
to the detuning, as can be found in Eqs.~(\ref{25})-(\ref{27}).
The plots in Fig.~\ref{fig5}(a) when the detuning is reduced to $\Delta=1.5\pi{\rm{GHz}}$ and $\Delta=1\pi{\rm{GHz}}$,
respectively, verify our deduction.
They clearly show the feasibility of the C-STIRSAP under the ``chosen path" method that
nearly perfect transfer efficiency can be obtained within a short operation
time while the pulses do not have to be strong.
For instance, we choose a point in this figure
when $\Delta=1.5\pi{\rm{GHz}}$ and $t_{\rm{f}}=1000 {\rm{ns}}$.
The corresponding time sequence of the four Rabi frequencies in Fig.~\ref{fig5}(b)
and the population evolution of the M-type system in Fig.~\ref{fig5}(c)
demonstrate the high transfer efficiency while the amplitudes of
$\tilde{\Omega}_{1}$ and $\tilde{\Omega}_{4}$ are just around $40 \pi {\rm{MHz}}$.

\begin{figure}[t]
\centering{\includegraphics[width=8.5cm]{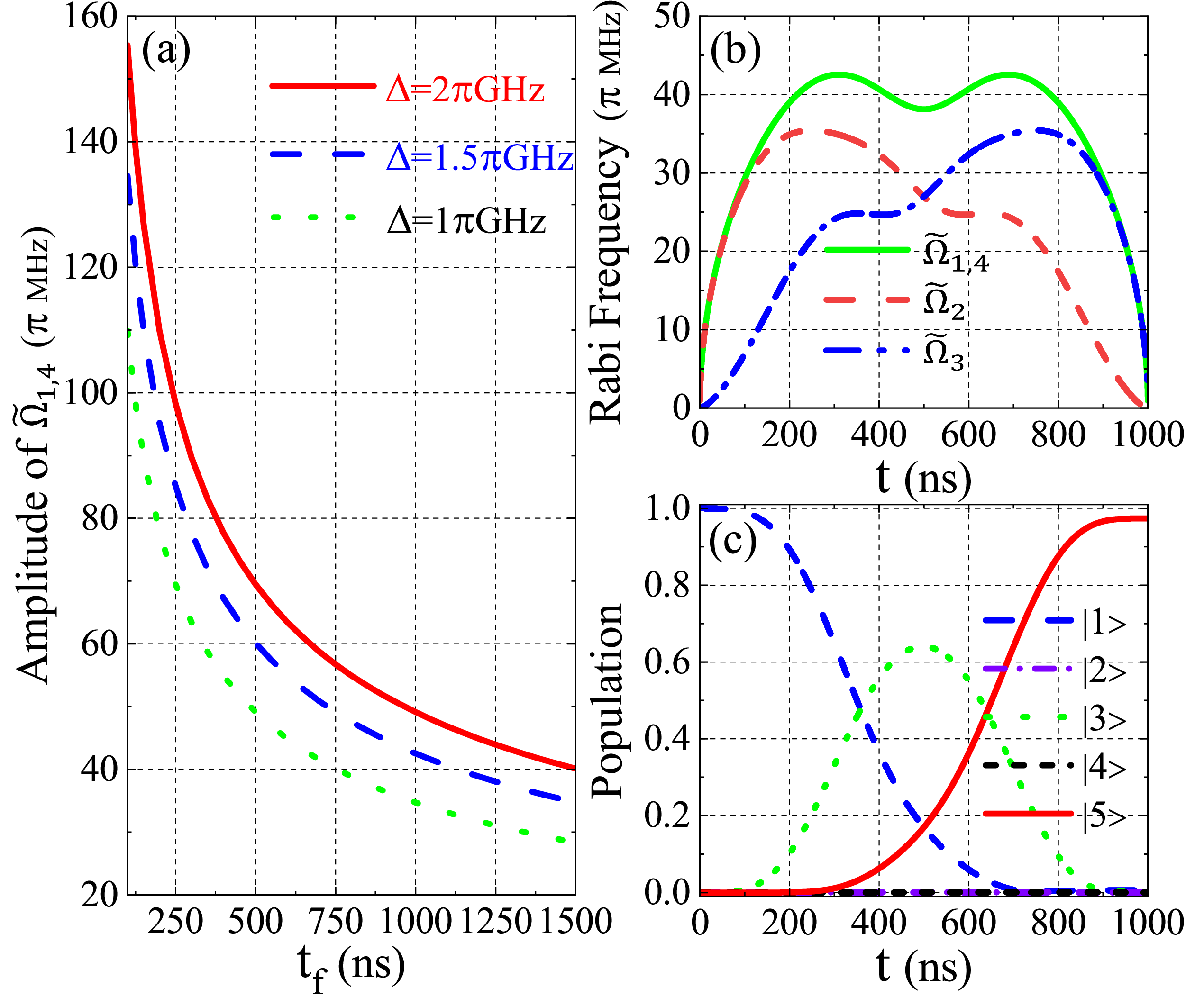}}
\caption{(a) Amplitudes of $\tilde{\Omega}_{1}$ and $\tilde{\Omega}_{4}$ as a function of the operation time
$t_{\rm{f}}$ at different detuning.
(b) and (c) present the time sequence of the four Rabi frequencies $\tilde{\Omega}_{j} (j=1,2,3,4)$
and the corresponding population transfer when the detuning and operation time are $\Delta=1.5\pi{\rm{GHz}}$
and $t_{\rm{f}}=1000 {\rm{ns}}$, respectively. Note the parameter $\gamma$ is
fixed at $0.3\pi$.}
\label{fig5}
\end{figure}

\begin{figure}[t]
\centering{\includegraphics[width=8cm]{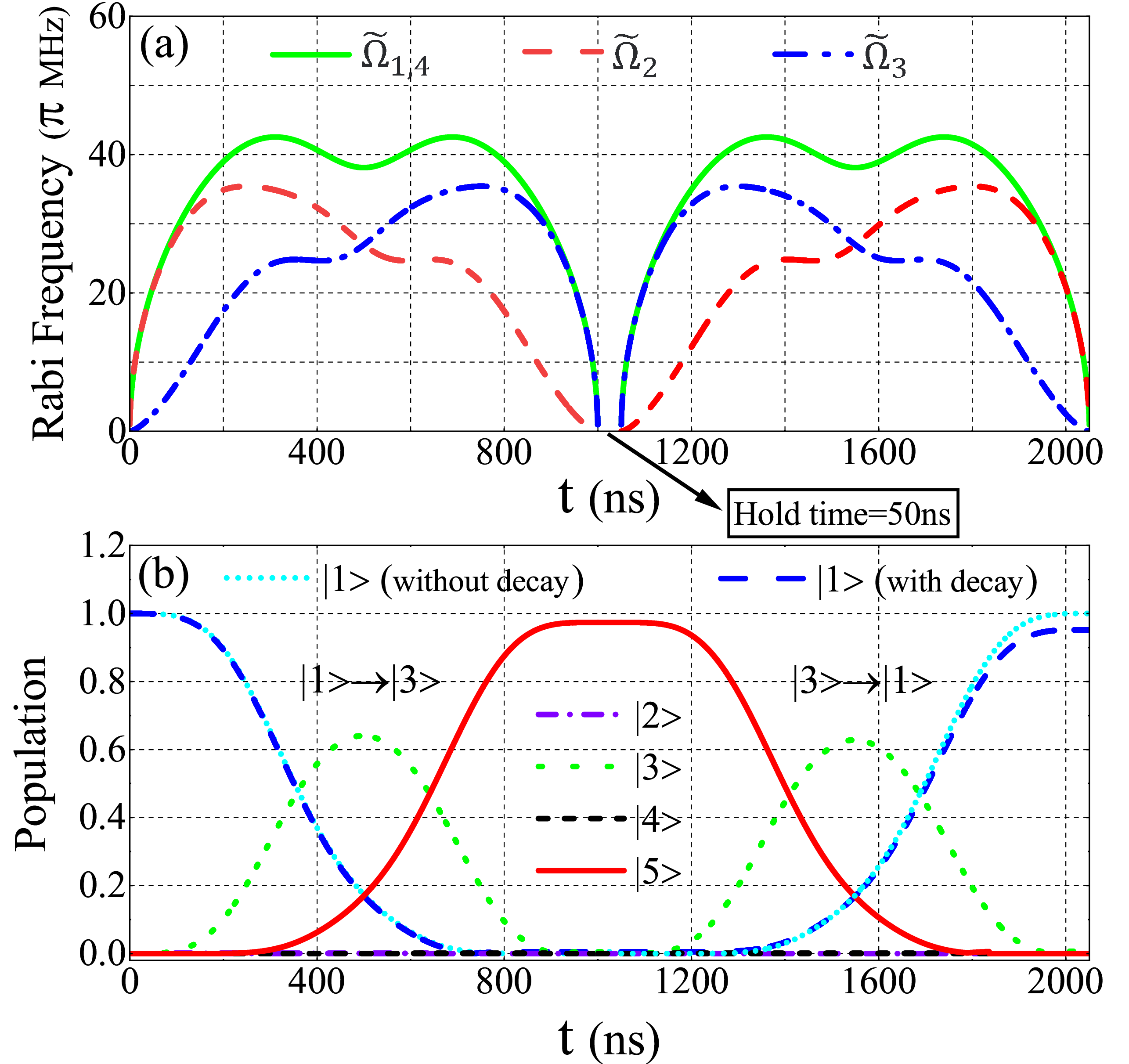}}
\caption{(a) Time sequence of the four Rabi frequencies $\tilde{\Omega}_{j} (j=1,2,3,4)$
for the creation and detection of ultracold deeply-bound molecules and
(b) the corresponding population transfer.
Population evolution of the Feshbach state $|1\rangle$ without considering the decay is also
presented as a comparison.
All parameters are the same as those in Fig.~\ref{fig5}(b) and (c).}
\label{fig6}
\end{figure}

Finally, we discuss briefly the detection of ultracold deeply-bound molecules.
High detection efficiency is important to test the created ultracold deeply-bound molecules,
it is also critical for the preparation and characterization of novel many-body phases of dipolar molecules,
such as crystalline bulk phases \cite{PhysRevLett.98.060404}, exotic density order \cite{PhysRevResearch.4.013235},
or spin order in optical lattices \cite{PhysRevLett.107.115301}.
Generally, we should first transfer the ultracold deeply-bound molecules back into the Feshbach molecules.
Then we can detect them by using the standard absorption imaging.
The back-transfer is usually realized by reversing the time sequence of the incident pulses
for the creation process.
Therefore, the detection efficiency critically depends on the creation efficiency of
ultracold deeply-bound molecules.
If the creation efficiency is $\chi$, then the detection efficiency will be $\chi^2$.
Here, we directly give an example of the detection following the creation results in
Figs.~\ref{fig5}(b) and~\ref{fig5}(c), which is presented in Fig.~\ref{fig6}.
In order to reverse the time sequence of the four Rabi frequencies
$\tilde{\Omega}_{j} (j=1,2,3,4)$,
the boundary condition of the parameter $\theta(t)$ is altered to be
$\theta(0)=\pi/2$ and $\theta(t_f)=\pi$. Therefore, $\theta(t)$ is redefined as
\begin{eqnarray}\label{29}
\theta(t)=&&\frac{\pi}{2}+\frac{\pi t}{2t_f}-\frac{1}{3}\sin(\frac{2\pi t}{t_f})+\frac{1}{24}\sin(\frac{4\pi t}{t_f}).
\end{eqnarray}

\noindent
As a comparison, we also give the population evolution of the Feshbach state $|1\rangle$ in the
ideal case of not considering the decay.
We can clearly see in Fig.~\ref{fig6} that there is always a high detection efficiency
under our C-STIRSAP protocol. The detection efficiency is about $95.16\%$, the result indicates a creation efficiency of about $97.55\%$. These results validate the potential of our approach for preserving the phase-space density of the ultracold mixture.

\section{\label{sec:level3}Conclusions}

To conclude, we have proposed a scheme of C-STIRSAP in M-type system
to create and detect ultracold deeply-bound molecules.
The C-STIRSAP can speed up the population transfer without using strong pulses, resulting in
the creation and detection with high efficiency.
The key of our protocol is reducing the M-type structure into a generalized $\Lambda$-type system
with the simplest resonant coupling, which
principally allows us to employ all kinds of methods in the frame of STA to accelerate the adiabatic procedure.
The simplification is realized under the assumption of AE together with a requirement of the relation among the four incident pulses.
Examples of using the counter-diabatic driving and ``chosen path" method verify the feasibility of the C-STIRSAP.
Since the cost of STA is generally the precisely-controlled incident pulses, we note that the requirement of the
relation among the four pulses in our C-STIRSAP protocol actually does not add extra complexity in experiments.
In the end, we believe that our protocol is a good supplement to the arsenal of STA in multi-state quantum systems,
which is of potential interest in applications where high-fidelity quantum control is essential (e.g.,
quantum information, atom optics and cavity QED).

\section*{Acknowledgments}

This work was supported by the National Natural Science Foundation of China (Grant Nos. 11974109, 12034007)
and the
Program of Shanghai Academic Research Leader (Grant No. 21XD1400700).

\bibliography{references}
\end{document}